\documentclass[apl,preprint,superscriptaddress,showpacs,floatfix]{revtex4}

\usepackage{graphicx}

\begin{document}

\title{Model for spin coupling disorder effects on the susceptibility of
antiferromagnetic nanochains}
\author{C.M. \surname {Chaves}}
\email{cmch@cbpf.br}
\affiliation{Centro Brasileiro de Pesquisas F\'\i sica, Rua Xavier Sigaud 150, Rio de
Janeiro, 22290-180, RJ, Brazil}
\author{ Thereza \surname{Paiva}}
\affiliation{Instituto de F\'\i sica, Universidade Federal do Rio de Janeiro, Caixa
Postal 68528, 21941-972, Rio de Janeiro RJ, Brazil}
\author{ J. d'Albuquerque \surname{e Castro}}
\affiliation{Instituto de F\'\i sica, Universidade Federal do Rio de Janeiro, Caixa
Postal 68528, 21941-972, Rio de Janeiro RJ, Brazil}
\author{Belita \surname {Koiller}}
\affiliation{Instituto de F\'\i sica, Universidade Federal do Rio de Janeiro, Caixa
Postal 68528, 21941-972, Rio de Janeiro RJ, Brazil}
\date{\today }

\begin{abstract}
The temperature dependence of the static magnetic susceptibility of
exchange-disordered antiferromagnetic Heisenberg spin-1/2 finite chains with
an odd number of spins is investigated as a function of size and type of
disorder in the exchange coupling. 
Two models for the exchange disorder distribution are considered. 
At sufficiently low temperatures each
chain behaves like an isolated spin-1/2 particle. As the size of the chains
increases, this analogy is lost and the chains evolve into the thermodynamic
limit behavior. The present study provides a simple criterion, based on
susceptibility measurements, to establish when odd-sized chains effectively
simulate a single spin-1/2 particle.
\end{abstract}

\maketitle



Precise placement of individual atoms in a host material would allow
improvements in the performance of currently available devices\cite%
{shanida05}, fabrication of new devices,\cite{Kane} as well as experimental
verification of basic properties predicted for low-dimensional systems.\cite%
{lounis} Control over atomic positioning on surfaces was first achieved over
a decade ago. More recently, significant progress has been reported in the
contexts of magnetic nanochains\cite{bru,hirji} and of donor placement in Si.%
\cite{oberbeck,shen,schofield} Accurate dopant positioning (e.g. P) in Si is
motivated not just by the trend dictated by Moore's law, requiring
fabrication of smaller and increasingly precise devices in Si, but also by
proposals for Si-based quantum computers,\cite{Kane} where the qubits involve
the electronic and nuclear spins of shallow donors. Spin 1/2 particles are
ideal candidates for qubits in Si because of their limited interactions with
their environment, leading to long coherence times.\cite{sarma} However, the
proposed one- and two-qubit gates, driven by electric and magnetic
fields, would have to be controlled within the length scale of a
single spin.\cite{Kane,friesen} An alternative definition of a qubit, for
which conditions on field control would be less severe, has been proposed by
Meier \textit{et al.}.\cite{meier1,meier2} They have investigated the
magnetic behavior of antiferromagnetic (AF) clusters of spin-1/2 particles,
which exhibit a $S^{z}=\pm 1/2$ doublet ground state\cite{bonner} and could
be used to define a logical qubit. Quantum gate operations would not affect
the coupling between spins within the chains, and would require the control
of electric and magnetic fields on the length scale of the spin array.

Linear arrays with odd number of P atoms in Si, in which the coupling
between the electronic spins on adjacent P donors is AF,\cite{Kane} meet in
principle the above conditions for defining qubits. However, the intensity
of such coupling is highly sensitive to the relative position os the donors.%
\cite{bk,hu} Changes in the donor positioning of just one lattice parameter
may alter the strength of the coupling by orders of magnitude. Therefore, at
the level of precision so far achieved in the techniques of sample
preparation, which is of about 1 nm,\cite{oberbeck,schofield} these chains
are bound to exhibit some degree of disorder in their structure, leading to
fluctuations in the exchange interaction $J$ between magnetic moments in the
chain. When the P atoms are positioned along a single [100] crystal axis,
with the inter-donor separations distributed around some target value, $J$
remains restricted to an interval around the average value $J_{0}$.\cite%
{sarma,hu} Assuming that fluctuations in the impurity positioning along the
[100] axis are of the order of 1 nm, the probability distribution of the
exchange interaction can be well described by a trimodal one, 
\begin{equation}
P_{tri}(J)=(1/3)\{\delta (J-J_{0})+\delta \lbrack J-(1+W)J_{0}]+\delta
\lbrack J-(1-W)J_{0}]\}~,
\end{equation}%
where $0<W<1$ is a parameter giving the degree of dispersion of $J$. On the
other hand, slight deviations on the P positioning, on the order of the interatomic distance with respect to the perfectly aligned chain along a [100] axis, lead to important differences. The distribution of values of $J$ turns out to be peaked near $J=0$ \cite{hu} and can be modeled by
an exponential one,%
\begin{equation}
P_{exp}(J)=\frac{1}{J_{0}}e^{-J/J_{0}}\,\Theta (J)~,
\end{equation}%
where $\,\Theta (J)$ is the unitary step function. In any case, the
occurrence of disorder in the exchange interaction within the cluster is a
relevant ingredient in determining its magnetic behavior. It is therefore a
key issue regarding the practical use of such clusters, in particular to
define qubits.

In the case of ordered linear chains of atomic spins
assembled on an insulating surface, scanning tunneling microscopy (STM)
measurements have been recently used to investigate the magnetic excitation
spectra of such systems, on the basis of which the strength of the coupling
between the spins could be assessed.\cite{hirji} However, for chains buried
in the host bulk material, as in the case of donor-based spin qubits,\cite%
{Kane} such an approach would not be applicable.

We show here that susceptibility measurements on samples
with chains of odd number of magnetic spin 1/2 particles constitute a
valuable tool for investigating their magnetic behavior, in particular to
determine the conditions under which these chains behave as an effective
spin 1/2 particle. 
Susceptibility measurements on a different nanoscale system, 
namely molecular nanomagnets, have proven useful to characterize their 
magnetic behavior.\cite{thomas}

We have carried out a detailed study of the temperature behavior of the
averaged susceptibility per spin, $\left\langle \chi _{N}\right\rangle $, of
AF spin-$1/2$ chains with odd number $N$ of spins. We have focused our
attention on two fabrication-related factors affecting the magnetic
properties of such chains, namely their length and type of disorder in the
exchange interaction. Both trimodal and exponential exchange distributions,
as given by Eqs. (1) and (2), respectively, have been considered. For the
former, $W$ was set equal to 0.5, which corresponds to a relatively wide
disorder distribution. For  $N\leq 3$ average susceptibilities have been
calculated analytically, whereas for larger sizes, quantum Monte Carlo 
simulations have been performed.\cite{ch} In what follows, susceptibilities
are given here in units of $\chi _{0}=g^{2}\mu _{B}^{2}/J_{0}$ and
temperature in units of $J_{0}/k_{B}$, where $\mu _{B}$ is the Bohr magneton
and $g(=2)$ is the Land\'{e} factor. In these units, the static
susceptibility of a single Heisenberg magnetic moment is given by $\chi
_{1}=S(S+1)/3T$, so that $T\chi _{1}=1/4$ when $S=1/2$.

\begin{figure}

\begin{center}

\resizebox{80mm} {!}{\includegraphics{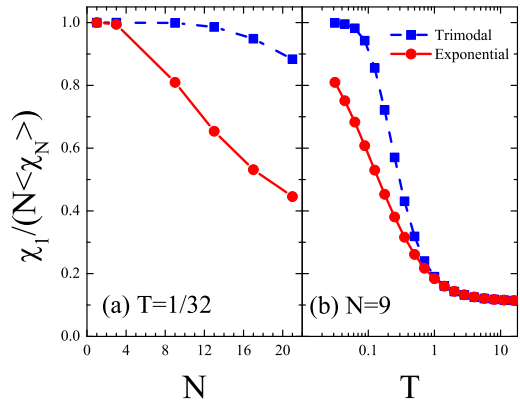}}
\caption{ \label{f2}(Color online) (a) Calculated susceptibility ratio, $\chi_{1}/(N\left\langle \chi _{N}\right\rangle)$  (single-spin susceptibility over the whole chain susceptibility) at low $T$ ($T=1/32$) versus $N$ for the trimodal and the exponential distributions and odd values of $N$ varying from 1 to 21.  (b) Susceptibility ratio versus $T$ calculated for $N=9$. 
}
\end{center}
\end{figure}

At low temperatures ($T<<1$), the behavior of $\left\langle \chi
_{N}\right\rangle $ should be analogous to that of a single $S=1/2$ spin. 
\cite{meier1,meier2} We have investigated the extent to which such analogy
holds under the combined effects of disorder, temperature and chain length $N
$. Fig.\ref{f2} (a) shows the susceptibility ratio $\chi _{1}/(N\langle \chi
_{N}\rangle )$ at $T=1/32$ for odd values of $N$ ranging from 1 to 17, and
for the two exchange distributions. Such ratio provides us with a simple 
quantitative criterion for assessing if the clusters behave collectively  
as a single spin: In this case the ratio should be 1. We notice that for the trimodal
distribution, the susceptibility ratio remains close to 1 over a relatively
wide range of values of $N$. However, in the case of exponential disorder, 
significant deviations from 1 rapidly occur as $N$ increases above 3. 
It is also interesting to look at the behavior of the susceptibility ratio as a
function of $T$, for fixed $N$. Results for $N=9$ are presented in Fig.\ref{f2} (b)
for the two exchange distributions. We clearly see that for the trimodal
distribution the ratio remains close to 1 over a temperature range wider
than the one corresponding to the exponential one. 
As $T$ increases, results for the two disorder distributions merge and the susceptibility ratio tends to $1/N$, 
indicating that in the high-$T$ limit the chain breaks into $N$ independent spins.
In all cases, the deviation of the susceptibility ratio from 1, for a given temperature and type of disorder, increases with $N$.
We conclude that using longer chains as qubits,\cite{meier1,meier2} with the advantage 
of facilitating qubit control by external fields, also brings more severe requirements in 
terms of achieving sufficiently low temperatures.

\begin{figure}

\begin{center}




\resizebox{80mm}{!}{\includegraphics{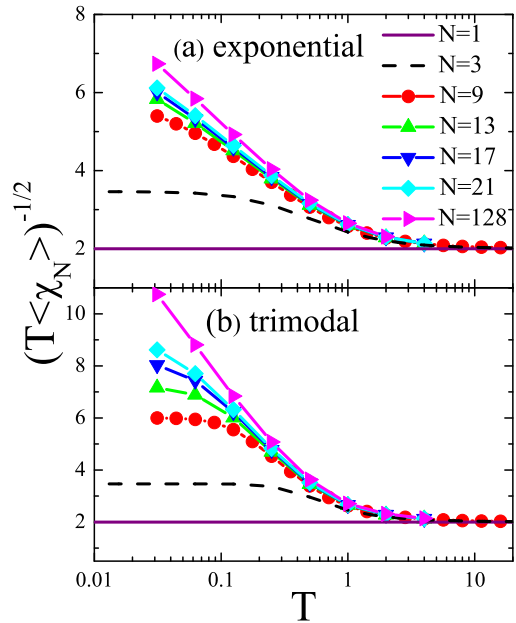}}
\caption{\label{f4}(Color online) Calculated $1/\sqrt{T\left\langle \chi _{N}\right\rangle}$ for the indicated values of $N$ for (a) exponential (b) trimodal  disorder distributions. Horizontal lines correspond to Curie-like behavior.}

\end{center}

\end{figure}

For both low and high temperature regimes, $\langle \chi _{N}\rangle $
exhibits a Currie-like behavior, though with different prefactors. In the
former, the coefficient is equal to $1/4N$, whereas in the latter, where
thermal fluctuations overcame the exchange interaction, it is equal to $1/4$. 
This can be clearly seen by plotting $(T\langle \chi _{N}\rangle )^{-1/2}$
as a function of $\log (T)$ over a rather wide range of temperatures, as in
Fig.\ref{f4}. In this sort of plotting, Curie-like $(\sim 1/T)$ behavior is
represented by a horizontal line. The figure shows results for different
values of $N$ and for exponential (a) and trimodal (b) distributions. In
both cases, we notice that in the intermediate temperature region, $%
(T\langle \chi _{N}\rangle )^{-1/2}$ shows a linear behavior with $\log (T)$%
, which becomes more pronounced as $N$ increases. We remark that such
behavior represents a precursor of the Fisher \cite{ma1,daniel} scaling-law,
according to which in the thermodynamic limit ($N\rightarrow \infty $) the
plot of $(T\langle \chi _{\infty }\rangle )^{-1/2}$ vs. $\log (T)$ results
in a straight line whose slope depends on just the exchange coupling
disorder distribution.\cite{ch,todo} Departure from the Fisher scaling
behavior below some characteristic temperature $T^{\ast }$ signals the onset
of the single spin behavior. Such temperature should provide an estimate for
the magnitude of the gap between the ground state doublet and the first
excited state. A plot of $\langle \chi _{N}\rangle $ as a function of $NT$
in a double-log scale for several odd values of $N$ is presented in Fig. \ref%
{f5}. We note that for $NT\lesssim 1$ all curves collapse onto a single one,
indicating the onset of the single spin behavior: It follows that $T^{\ast
}\sim 1/N$, supporting our interpretation in terms relating $T^{\ast }$ to
the first excitation gap, which is known to scale with the inverse of the
number of sites in the chain.\cite{meier2}

In conclusion, we have studied the temperature behavior of the magnetic
susceptibility of odd-numbered linear spin chains. Our calculations
encompass a wide range of values of $N$ and of $T$, and clearly establish
the existence of three temperature regions where the spin susceptibility of
the chains exhibit distinct behaviors. Disorder in the intra-chain exchange
couplings plays an important role in our results. For a disorder
distribution that does not include extremely small values of $J$, the spin
cluster analogy with single spin 1/2 particles remains robust at low-$T$,
even if the distribution is considerably wide, as in the case of the
trimodal distribution considered here. However, if arbitrarily small values
of $J$ occur, which is probably unavoidable under current samples
fabrication capabilities, restricting the system dynamics to the doublet
ground state manifold would require unrealistically low operation
temperatures. We recall that in the case of donor-based spin qubits, sample
preparation should include an overgrowth stage after atomic positioning at a
surface. We have also determined how the temperature below which the cluster
behaves as a single spin, $T^{\ast }$, scales with $N$, establishing its
relation with the first excitation gap. The present work sheds light on
relevant points regarding the magnetic behavior of antiferromagnetic
nanochains and its perspective for application as qubits in spin-based
solid-state devices.

\begin{figure}

\begin{center}




\resizebox{95mm}{!}{\includegraphics{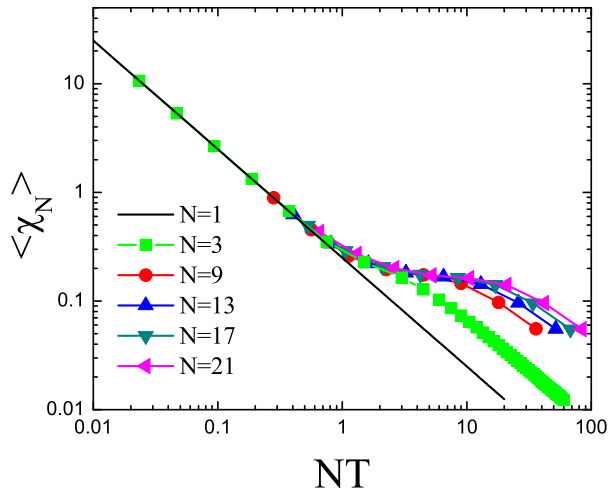}}
\caption{\label{f5}(Color online) Average susceptibility for odd number of spins chains versus $NT$ in the case of trimodal disorder. The data suggest that {$T^*(N)\sim0.8/N$} for this particular type of disorder.}

\end{center}

\end{figure}
\begin{acknowledgments}
We thank R. T. Scalettar, F. H\'ebert and G. G. Batrouni for valuable
discussions. This work has been partially supported by the Brazilian
agencies PCI/MCT, CNPq, FAPERJ and the Millennium Institute of
Nanotechnology.
\end{acknowledgments}

\nobreak

\end{document}